\documentstyle[preprint,aps,epsfig]{revtex}
\begin{document}

\def\ibid#1#2#3{{\it ibid.} {\bf #1} (#2) #3}
\def\epjc#1#2#3{Eur. Phys. J. C {\bf #1} (#2) #3}
\def\ijmpa#1#2#3{Int. J. Mod. Phys. A {\bf #1} (#2) #3}
\def\mpl#1#2#3{Mod. Phys. Lett. A {\bf #1} (#2) #3}
\def\npb#1#2#3{Nucl. Phys. {\bf B#1} (#2) #3}
\def\plb#1#2#3{Phys. Lett. B {\bf #1} (#2) #3}
\def\prd#1#2#3{Phys. Rev. D {\bf #1} (#2) #3}
\def\prl#1#2#3{Phys. Rev. Lett. {\bf #1} (#2) #3}
\def\rep#1#2#3{Phys. Rep. {\bf #1} (#2) #3}
\def\zpc#1#2#3{Z. Phys. {\bf #1} (#2) #3}

\title{CP violating transverse lepton polarizaion in $B\to D^{(*)}\ell\bar\nu$
including tensor interactions}
\author{
  Jong-Phil Lee\footnote{Email address:~jplee@phya.yonsei.ac.kr}}
\address{Institute of Physics and Applied Physics, Yonsei University, Seoul, 120-749, Korea}
\maketitle
\begin{abstract}
We give a model-independent analysis of CP violating lepton polarization in the
exclusive semileptonic $B$ decay of $B\to D^{(*)}\ell\bar\nu$ including 
dimension six four-fermion tensor interactions at the heavy quark limit.
It is shown that the tensor interactions should not be neglected if the 
associated couplings are comparable to others.
The effect of tensor interactions on the transverse lepton polarization appears 
more dramatically in $B\to D$ than in $B\to D^*$.
In the leptoquark model, the average transverse lepton polarization is 
estimated to be
$|\overline{P^\perp_D}|\simeq 0.26$ and
$|\overline{P^\perp_{D^*}}|\simeq 0.076$ with commonly used model parameters.

\end{abstract}
\section{Introduction}

CP violation (CPV) is one of the most important puzzles in particle physics.
The origin of it still remains as a great mistery.
In the standard model (SM), only one Cabibbo-Kobayashi-Maskawa (CKM) \cite{CKM}
phase explains CP violation.
Although it describes CPV successfully in $K$-$\bar{K}$ system, 
only one CKM phase is too few to explain various possibilities of CP violation.
It is nowadays not unnatuall to think of the new physics beyond the SM.
Various kinds of extensions of the SM contain the CPV phases and
may contribute greatly to the CPV observables \cite{Worah}.
The study of CPV thus not only provides a deeper understanding of the CKM
structure but also gives some clues of the new physics such as supersymmetry
\cite{Ali,Gabrielli,Masiero} or the minimal flavor violating extension of the SM
\cite{Ciuchini,Buras,Lunghi}.
\par
With the beginning of the $B$ factory era \cite{Bfactory}, a lot of data are 
accumulating in BABAR and Belle \cite{sin2beta}.
The measured value of $\sin 2\beta$ strongly suggests that CP is broken, and
not an approximate symmetry.
However, the measured values of CKM parameters reside within the scope of the
SM.
No crucial evidences for new physics are reported yet.
In probing a new physics beyond the SM,  CP-odd observables which do not appear
in the SM prediction are particularly interesting.
\par
In this paper we give a model-independent analysis of transverse lepton 
polarization to the decay plane in exclusive semileptonic $B$ decays
$B\to D^{(*)}\ell\bar{\nu}$.
Transverse lepton polarization is a triple-vector-correlations given by
$\vec{s}_\ell\cdot(\vec{p}_{D^{(*)}}\times\vec{p}_\ell)$
where $\vec{s}_\ell$ is the spin vector of lepton, $\vec{p}_{D^{(*)}}$ is the
momentum of $D^{(*)}$, and $\vec{p}_\ell$ is the momentum of lepton.
This quantity is CP odd just as the transverse $\mu$ polarization in $K_{\mu 3}$
decay \cite{Sakurai}.
In general this transverse lepton polarization is proportional to the imaginary
part of multiples of hadronic form factors \cite{Belanger}.
But the hadronic form factors are real in SM; 
the SM predicts no transverse lepton polarizations.
The observation of nonzero transverse lepton
polarization is therefore a signal of a new physics beyond the SM 
\cite{Koerner,Garisto,Wu}.
We consider all the possible dimension six four-fermion interactions 
\cite{Goldberger}.
In the previous work of \cite{Wu}, it is shown that $B\to D\ell\bar{\nu}$ is 
sensitive to the new scalar interactions while $B\to D^*\ell\bar{\nu}$ to the 
new pseudoscalar interactions.
In this work, special attentions are paid to the tensor interactions 
to see their effects.
Models such as leptoquarks can have a sizable tensor contributions.
\par
The main source of theoretical uncertainties in the analysis of semileptonic
$B$ decays is the hadronic matrix elements.
In the SM, $B\to D\ell\bar{\nu}$ involves two hadronic form factors while
$B\to D^*\ell\bar{\nu}$ does four.
Thanks to the heavy quark effective theory (HQET), all these form factors
are related to one universal Isgur-Wise (IW) function in $m_Q\to\infty$ limit 
where $m_Q$ is the heavy quark mass \cite{Isgur}.
We work in the heavy quark limit for simplicity.
However, it is inevitable to use nonperturbative methods for a complete 
analysis.
We adopt the results from QCD sum rule calculations.
\par
In the next section, the interaction Lagrangian 
containing general four-fermion interactions is given and their 
contributions to the SM form factors are described.
In Sec.\ III transverse lepton polarization is calculated to see the effect of
tensor interactions. 
Sec.\ IV contains the results and discussions.
As an application of the results, the leptoquark model is considered.
The summary is given in Sec.\ V.

\section{Interaction Lagrangian and form factors}

Semileptonic $B$ decays are well described in the SM by the following 
interaction lagrangian:
\begin{equation}
{\cal L}_{\rm SM}
=-\frac{G_F}{\sqrt{2}}V_{cb}\bar{c}\gamma_\mu
 (1-\gamma_5)b\bar{l}\gamma^\mu(1-\gamma_5)\nu + {\rm h.c.}~,
\label{sm}
\end{equation}
where $G_F$ is the Fermi constant and $V_{cb}$ is the CKM matrix element.
The effects of new physics can be parametrized in a similar manner by extending
the coupling and the $V-A$ structure as \cite{Goldberger}
\begin{equation}
{\cal L}_{\rm new}=
\frac{G_F}{\sqrt{2}}V_{cb}\sum_{p,q,r}g^{p}_{q,r}
  \bar{c}\Gamma^p b_q\bar{l}_r\Gamma^p\nu+{\rm h.c.}~,
\label{new}
\end{equation}
where $\frac{G_F}{\sqrt{2}}V_{cb}g^{p}_{q,r}$ are the new couplings and
$p$ runs over 
\begin{equation}
p=S({\rm Scalar})~,~~V({\rm Vector})~,~~T({\rm Tensor})~.
\end{equation}
Note that $g^{p}_{q,r}$ are dimensionless and they can be complex in general.
$\Gamma^p$ is the corresponding $\gamma$ matrices like
\begin{equation}
\Gamma^S=1~,~~(\Gamma^V)^\mu=\gamma^\mu~,
~~(\Gamma^T)^{\mu\nu}=\sigma^{\mu\nu}=\frac{i}{2}[\gamma^\mu,\gamma^\nu]~,
\end{equation}
where $q$ and $r$ represent the helicity of $b$-quark and lepton $l$ respectively, 
so they are $L$ (left-handed) or $R$ (right-handed).
In Eq.(\ref{new}), right-handed neutrinos are also considered. 
The relevant coupling constants for the right-handed neutrinos are
$g^S_{\mu L}$, $g^V_{\mu R}$, $g^T_{\mu L}$.
The right-handed neutrino contributes to the squared matrix element only
at the order of ${\cal O}(g^2)$, so we neglect the four-fermion operators
involving a right-handed neutrino from now on.

\par
The hadronic matrix elements are specified by the two form factors for
$B\to D l \bar{\nu}$ and four form factors for $B\to D^* l\bar{\nu}$
as follows:
\begin{eqnarray}
\langle D(p^\prime)|\bar{c}\gamma^\mu b|B(p)\rangle
&=&f_+(p+p^\prime)^\mu+f_-(p-p^\prime)^\mu ~,\nonumber\\
\langle D^*(p^\prime,\epsilon)|\bar{c}\gamma^\mu b|B(p)\rangle
&=&i\frac{F_V}{m_B}\epsilon^{\mu\nu\alpha\beta}\epsilon^*_\nu
   (p+p^\prime)_\alpha q_\beta~,\nonumber\\
\langle D^*(p^\prime,\epsilon)|\bar{c}\gamma^\mu\gamma_5 b|B(p)\rangle
&=&-F_{A0}m_B\epsilon^{*\mu}
   -\frac{F_{A+}}{m_B}(p+p^\prime)^\mu\epsilon^*\cdot q
   -\frac{F_{A-}}{m_B}q^\mu\epsilon^*\cdot q~,
\label{matrix}
\end{eqnarray}
where $p$ and $p^\prime$ are the four-momenta of the $B$ and $D^{(*)}$,
respectively, $\epsilon$ is the polarization vector of $D^*$, 
and $q=p-p^\prime$.
For the case of $B\to D l\bar{\nu}$,
$\langle D(p^\prime)|\bar{c}\gamma_\mu\gamma_5b|B(p)\rangle=0$
because there is no way to construct axial vector using only $p$ and
$p^\prime$.
Another kinds of hadronic matrix elements are obtained from 
Eq.(\ref{matrix}) by using the Dirac equations
\begin{eqnarray}
\langle D(p^\prime)|\bar{c}b|B(p)\rangle
&=&\frac{m_B^2}{m_b-m_c}\Bigg[f_+(1-r_D)+f_-\frac{q^2}{m_B^2}\Bigg]
\nonumber\\
\langle D(p^\prime)|\bar{c}\gamma_5b|B(p)\rangle&=&0\nonumber\\
\langle D^*(p^\prime,\epsilon)|\bar{c}b|B(p)\rangle&=&0\nonumber\\
\langle D^*(p^\prime,\epsilon)|\bar{c}\gamma_5b|B(p)\rangle
&=&\frac{m_B}{m_b+m_c}\epsilon^*\cdot q\Bigg[
   F_{A0}+F_{A+}(1-r_{D^*})+F_{A-}\frac{q^2}{m_B^2}\Bigg]~,
\end{eqnarray}
where $m_b$ and $m_c$ are the $b$ and $c$ quark masses respectively, 
$r_D=m_D^2/m_B^2$ and $r_{D^*}=m_{D^*}^2/m_B^2$.
The tensor quark bilinear can also be written as above, quite easily
in the case of $B\to D l\bar{\nu}$,
\begin{eqnarray}
\langle D(p^\prime)|\bar{c}\sigma_{\mu\nu}b|B(p)\rangle
&=&\frac{m_B^2}{m_b-m_c}\Bigg[f_+(1-r_D)+f_-\frac{q^2}{m_B^2}\Bigg]
\frac{i(p\cdot p^\prime-m_bm_c)}{m_B^2m_D^2-(p\cdot p^\prime)^2}
(p_\mu p^\prime_\nu-p_\nu p^\prime_\mu)~,\nonumber\\
\langle D(p^\prime)|\bar{c}\sigma_{\mu\nu}\gamma_5b|B(p)\rangle
&=&\frac{m_B^2}{m_b-m_c}\Bigg[f_+(1-r_D)+f_-\frac{q^2}{m_B^2}\Bigg]
\frac{(m_bm_c-p\cdot p^\prime)}{m_B^2m_D^2-(p\cdot p^\prime)^2}
\epsilon_{\mu\nu\alpha\beta}p^\alpha p^{\prime\beta}~.
\end{eqnarray}

\par
For the hadronic matrix element of $B\to D^* l\bar{\nu}$,
we find that it is very convenient to use HQET.
In HQET, there is a symmetry of heavy quark spin and flavor in 
$m_Q\to\infty$ limit where $m_Q$ is the heavy quark mass.
Introducing interpolating fields for the description of heavy mesons,
the most general form of the matrix elements can be calculated as
\cite{Wise}
\begin{eqnarray}
\langle P_{Q_j}(v^\prime)|\bar{h}^{(j)}_{v^\prime}\Gamma h^{(i)}_v
|P_{Q_i}(v)\rangle &\propto&
\xi(v\cdot v^\prime){\rm Tr}\Bigg[
  \gamma_5\Big(\frac{v^\prime \hspace{-3mm}/+1}{2}\Big)\Gamma
  \Big(\frac{v \hspace{-2mm}/+1}{2}\Big)\gamma_5\Bigg]~,\nonumber\\
\langle P_{Q_j}^*(v^\prime,\epsilon)|\bar{h}^{(j)}_{v^\prime}\Gamma 
h^{(i)}_v |P_{Q_i}(v)\rangle &\propto&
\xi(v\cdot v^\prime){\rm Tr}\Bigg[
  \epsilon^*\hspace{-3mm}/\Big(\frac{v^\prime \hspace{-3mm}/+1}{2}\Big)\Gamma
    \Big(\frac{v \hspace{-2mm}/+1}{2}\Big)\gamma_5\Bigg]~,
\end{eqnarray}
where $P_{Q_i}^{(*)}(v)$ is the heavy meson state of four-velocity $v$ with
a heavy quark $Q_i$, $h_v^i$ is the heavy quark field, $\Gamma$ is any kind
of $\gamma$ matrices, and $\xi(v\cdot v^\prime)$ is the Isgur-Wise function.
Using the above expressions,
\begin{eqnarray}
\langle D^*(p^\prime,\epsilon)|\bar{c}\sigma_{\mu\nu}b|B(p)\rangle&=&
F_T\epsilon_{\mu\nu\alpha\beta}\big(\epsilon^{*\alpha}p^\beta
  +\frac{1}{\sqrt{r_{D^*}}}\epsilon^{*\alpha}p^{\prime\beta}\big)~,\nonumber\\
\langle D^*(p^\prime,\epsilon)|\bar{c}\sigma_{\mu\nu}\gamma_5b|B(p)\rangle&=&
-iF_T\Big[\epsilon^*_\mu p_\nu-\epsilon^*_\nu p_\mu+\frac{1}{\sqrt{r_{D^*}}}
  (\epsilon^*_\mu p^\prime_\nu-\epsilon^*_\nu p^\prime_\mu)\Big],
\end{eqnarray}
where
\begin{equation}
F_T=-\frac{m_B m_c}{p\cdot p^\prime+m_B m_{D^*}}\Big[F_{A0}-
    \{(p\cdot p^\prime)^2/m_{D^*}^2-1\}(F_{A+}+F_{A-})\Big]~.
\label{FT}
\end{equation}
In the heavy quark limit where $m_Q\to \infty$,
not all the form factors are independent, but
they are related to the IW function $\xi(v\cdot v^\prime)$,
\begin{eqnarray}
f_\pm&=&\pm\frac{1\pm\sqrt{r_D}}{2\sqrt[4]{r_D}}\xi(w)~,\nonumber\\
F_V=F_{A+}&=&-F_{A-}=\frac{1}{2\sqrt[4]{r_{D^*}}}\xi(w)~,\nonumber\\
F_{A0}&=&-\sqrt[4]{r_{D^*}}(w+1)\xi(w)~,
\end{eqnarray}
where $w=v\cdot v^\prime=(m_B^2+m_{D^{(*)}}^2-q^2)/(2m_B m_{D^{(*)}})$.
With these hadronic matrix elements, the effects of ${\cal L}_{\rm new}$ in
Eq.\ (\ref{new}) appear as a slight modification of the form factors:
\begin{eqnarray}
f_+&\to&f^\prime_+=f_+(1+\delta_++\Delta_+)~,\\
f_-&\to&f^\prime_-=f_-(1+\delta_-+\Delta_-)~,\\
F_V&\to&F^\prime_V=F_V(1+\delta_V+\Delta_V)~,\\
F_{A0}&\to&F^\prime_{A0}=F_{A0}(1+\delta_{A0}+\Delta_{A0})~,\\
F_{A+}&\to&F^\prime_{A+}=F_{A+}(1+\delta_{A+}+\Delta_{A+})~,\\
F_{A-}&\to&F^\prime_{A-}=F_{A-}(1+\delta_{A-}+\Delta_{A-})~,
\end{eqnarray}
where
\begin{eqnarray}
\delta_+&=&-G_V~,\\
\delta_-&=&-G_V-\frac{G_S}{m_\ell}\frac{m_B^2}{m_b-m_c}\Big[\frac{f_+}{f_-}
          (1-r_D)+\frac{q^2}{m_B^2}\Big]~,\\
\delta_V&=&-G_V~,\\
\delta_{A0}&=&G_A~,\\
\delta_{A+}&=&G_A~,\\
\delta_{A-}&=&G_A-\frac{G_P}{m_\ell}\frac{m_B^2}{m_b+m_c}\Big[
             \frac{F_{A0}}{F_{A-}}+\frac{F_{A0}}{F_{A-}}(1-r_{D^*})
	     +\frac{q^2}{m_B^2}\Big]~,\\
\Delta_+&=&-2g^T_{RR}m_\ell
  \frac{p\cdot p^\prime-m_bm_c}{(p\cdot p^\prime)^2-m_B^2m_D^2}
  \frac{m_B^2}{m_b-m_c}\Big[1-r_D+\frac{f_-}{f_+}\frac{q^2}{m_B^2}\Big]~,\\
\Delta_-&=&-4\frac{g^T_{RR}}{m_\ell}
  \frac{p\cdot p^\prime-m_bm_c}{(p\cdot p^\prime)^2-m_B^2m_D^2}
  \frac{m_B^2}{m_b-m_c}\Big[\frac{f_+}{f_-}(1-r_D)+\frac{q^2}{m_B^2}\Big]
  \nonumber\\
  &&\times(p\cdot p^\prime-m_B^2+2p\cdot p_\nu+m^2_\ell/2)~,\\
\Delta_V&=&-2g^T_{RR}\frac{F_T}{F_V}\frac{m_B}{m_\ell}(1+1/\sqrt{r_{D^*}})~,\\
\Delta_{A0}&=&4g^T_{RR}\frac{F_T}{F_{A0}}\Big[(1+1/\sqrt{r_{D^*}})
  \frac{p^\prime\cdot p_\nu-p^\prime\cdot p_l}{m_B m_\ell}-m_\ell/m_B\Big]~,\\
\Delta_{A+}&=&-2g^T_{RR}\frac{F_T}{F_{A+}}(1+1/\sqrt{r_{D^*}})\frac{m_B}{m_\ell}
  \frac{\epsilon^*\cdot(p_\nu-p_l)}{\epsilon^*\cdot q}~,\\
\Delta_{A-}&=&2g^T_{RR}\frac{F_T}{F_{A-}}\frac{m_B}{m_\ell}\Big[
  (1+1/\sqrt{r_{D^*}})
  \frac{\epsilon^*\cdot(p_\nu-p_l)}{\epsilon^*\cdot q}
  +4\frac{\epsilon^*\cdot p_l}{\epsilon^*\cdot q}-2\Big]~,
\end{eqnarray}
and $G_V=g^V_{LL}+g^V_{RL}$, $G_A=-g^V_{LL}+g^V_{RL}$, $G_S=g^S_{LR}+g^S_{RR}$,
$G_P=-g^S_{LR}+g^S_{RR}$.
Here the terms of $\Delta$ are the corrections due to the tensor interactions.
In either case of $B\to D^{(*)}\ell\bar{\nu}$, as can be seen in the above 
expressions, the tensor interaction only contributes through $g^T_{RR}$.
This means that when tensor interaction is considered, only the right-handed 
$b$-quark spinor involves.

\section{Transverse lepton polarization in $B\to D^{(*)} \ell \bar\nu$}

As discussed in the introduction, the transverse component of lepton
polarization to the decay plane is CP-odd observable.
This transverse polarization of lepton can easily be obtained from the decay
amplitude using the spin projection operator $(1+\gamma_5 s\hspace{-2mm}/)/2$.
The transverse polarization of lepton is
\begin{equation}
P_{D^{(*)}}^\perp=\frac{|{\cal M}(D^{(*)},\vec{n})|^2
                        -|{\cal M}(D^{(*)},-\vec{n})|^2}
          {|{\cal M}(D^{(*)},\vec{n})|^2
	  +|{\cal M}(D^{(*)},-\vec{n})|^2}~,
\end{equation}
where $\vec{n}=(\vec{p}_D\times\vec{p}_\ell)/|\vec{p}_D\times\vec{p}_\ell|$,
and ${\cal M}(\pm\vec{n})$ is the decay amplitude with the lepton spin vector
along $\pm\vec{n}$.
The decay amplitudes are given by
\begin{eqnarray}
{\cal M}(D)&=&
-\frac{G_F}{\sqrt{2}}V_{cb}\bar{\ell}(p_\ell)\gamma^\mu(1-\gamma_5)\nu(p_\nu)
[f^\prime_+(p+p^\prime)_\mu+f^\prime_-(p-p^\prime)_\mu]~,\\
{\cal M}(D^*)&=&
-\frac{G_F}{\sqrt{2}}V_{cb}\bar{l}(p_\ell)\gamma^\mu(1-\gamma_5)\nu(p_\nu)
\epsilon^*_\rho{\cal M}^{\rho\mu}~,\nonumber\\
{\cal M}^{\rho\mu}&=&
i\frac{F^\prime_V}{m_B}\epsilon^{\mu\rho\alpha\beta}(p+p^\prime)_\alpha q_\beta
+F^\prime_{A0}m_B g^{\mu\rho}
+\frac{F_{A+}(1+\delta_{A+})}{m_B}(p+p^\prime)^\mu q^\rho  \nonumber\\
&&+\Bigg[\frac{F_{A-}(1+\delta_{A-})}{m_B}-\frac{4g^T_{RR}F^T}{m_\ell}\Bigg]
  q^\mu q^\rho
+\frac{2g^T_{RR}F^T}{m_\ell}(1+1/\sqrt{r_{D^*}})(p+p^\prime)^\mu
  (p_l-p_\nu)^\rho   \nonumber\\
&&+\frac{2g^T_{RR}F^T}{m_\ell}\Bigg[
  (1+1/\sqrt{r_{D^*}})(p_\nu-p_\ell)^\rho+4p_l^\rho\Bigg]q^\mu~,
\end{eqnarray}
where we have extracted out $\epsilon^*_\rho$ explicitly in ${\cal M}(D^*)$
to use 
$\sum_{\rm pol.}\epsilon_\mu\epsilon^*_\nu
  =-g_{\mu\nu}+p^\prime_\mu p^\prime_\nu/m_{D^*}^2$.

\par
Now the transverse lepton polarization in $B\to D \ell\bar{\nu}$ is given by
\begin{equation}
P^\perp_D=-\lambda_D(x,y){\rm Im}(2f^\prime_+f^{\prime *}_-)~,
\label{pol1}
\end{equation}
with
\begin{equation}
\lambda_D(x,y)=\frac{\sqrt{r_\ell}}{\rho_D(x,y)}
 \sqrt{(x^2-4r_D)(y^2-4r_\ell)-4\Bigg(1-x-y+\frac{1}{2}xy+r_D+r_\ell\Bigg)^2}~,\\
 \label{lam1}
\end{equation}
\begin{equation}
\rho_D(x,y)=|f^\prime_+|^2g_1(x,y)
  +2{\rm Re}(f^\prime_+ f^{\prime *}_-)g_2(x,y)+|f^\prime_-|^2g_3(x)~,
\end{equation}
where $r_\ell=m_\ell^2/m_B^2$, $x=2p\cdot p^\prime/p^2=2E_D/m_B$ and 
$y=2p\cdot p_l/p^2=2E_l/m_B$ in $B$ rest frame.
The kinematic functions $g_i(x,y)$ are given in the appendix.
Here $\rho_D(x,y)$ is proportional to the partial decay rate as
\begin{equation}
\frac{d^2\Gamma(B\to D \ell\bar{\nu})}{dxdy}
= \frac{G_F^2|V_{cb}|^2m_B^5}{128\pi^3}\rho_D(x,y)~.
\end{equation}
In Eq.(\ref{pol1}), ${\rm Im}(f^\prime_+f^{\prime *}_-)$ can have a finite
value other than zero 
because the new couplings are complex in general.
More explicitly,
\begin{eqnarray}
{\rm Im}(f^\prime_+f^{\prime *}_-)&=&
\frac{f_+}{\sqrt{r_\ell}(\sqrt{r_b}-\sqrt{r_c})}
[f_+(1-r_D)+f_-(1+r_D-x)]\nonumber\\
&&\times{\rm Im}\Bigg[G_S
+4g^T_{RR}\frac{x-2\sqrt{r_br_c}}{4r_D-x^2}\Big\{
  r_\ell(f_-/f_+-1)-2+x+2y\Big\}\Bigg]~,
\label{im}
\end{eqnarray}
where $r_{b(c)}=m_{b(c)}^2/m_B^2$.
The first term in Eq.(\ref{im}) is the same as Eq.(31) of \cite{Wu}, while
the second term represents the contribution of tensor interactions.
\par
The transverse lepton polarization in $B\to D^* \ell\bar{\nu}$ is quite similar
in form to that of $B\to D\ell\bar{\nu}$.
It is given by
\begin{eqnarray}
P^\perp_{D^*}&=&-\lambda_{D^*}(x,y)\Bigg[
  {\rm Im}(F^\prime_{A0}{\tilde F}^{*}_{A+})\Big(\frac{x}{2r_{D^*}}+1\Big)
 +{\rm Im}(F^\prime_{A0}{\tilde F}^{\prime*}_{A-})\Big(\frac{x}{2r_{D^*}}-1\Big)
\nonumber\\
&& 
 +{\rm Im}({\tilde F}_{A+}{\tilde F}^{\prime*}_{A-})
   \Big(\frac{x^2}{2r_{D^*}}-2\Big)
 +4{\rm Im}(F^\prime_{A0}g^{T*}_{RR}F^*_T)
  \frac{\sqrt{r_{D^*}}-r_\ell+x+y-1}{r_{D^*}\sqrt{r_\ell}}\nonumber\\
&&
 +{\rm Im}\Big\{({\tilde F}_{A+}+{\tilde F}^\prime_{A-})g^{T*}_{RR}F^*_T\Big\}
  (1+1/\sqrt{r_{D^*}})\frac{(x+y-2)(2r_{D^*}-x)+2r_\ell x}{r_{D^*}\sqrt{r_\ell}}
\nonumber\\&&
 +4{\rm Im}({\tilde F}_{A+}g^{T*}_{RR}F^*_T)\frac{x^2}{r_{D^*}\sqrt{r_\ell}}
\nonumber\\&&
 +8{\rm Im}(F^\prime_V g^{T*}_{RR}F^*_T)\frac{1}{\sqrt{r_\ell}}
  \Big\{2-x-y+\frac{1}{\sqrt{r_{D^*}}}(1-y+r_\ell-r_{D^*})\Big\}
 \Bigg]~,
\label{pol2}
\end{eqnarray}
where 
${\tilde F}_{A+}=F_{A+}(1+\delta_{A+})$, 
${\tilde F}^\prime_{A-}=F_{A-}(1+\delta_{A-})-\frac{4}{\sqrt{r_\ell}}g^T_{RR}F_T$,
$x=2p\cdot p^\prime/p^2=2E_{D^*}/m_B$, and
\begin{equation}
\lambda_{D^*}=
 \frac{\sqrt{r_\ell}}{\rho_{D^*}(x,y)}
  \sqrt{(x^2-4r_{D^*})(y^2-4r_\ell)-4\Bigg(1-x-y+\frac{1}{2}xy+r_{D^*}+r_\ell\Bigg)^2}~,\\
  \label{lam2}
\end{equation}
\begin{eqnarray}
\rho_{D^*}(x,y)&=&
|F_{A0}^\prime|^2f_1(x,y)+|{\tilde F}_{A+}|^2f_2(x,y)
+|{\tilde F}_{A-}^\prime|^2f_1(x,y)+|F_{V}^\prime|^2f_4(x,y)\nonumber\\&&
+2{\rm Re}(F_{A0}^\prime {\tilde F}_{A+}^*)f_5(x,y)
+2{\rm Re}(F_{A0}^\prime {\tilde F}_{A-}^{\prime*})f_6(x,y)\nonumber\\&&
+2{\rm Re}({\tilde F}_{A+} {\tilde F}_{A-}^{\prime *})f_7(x,y)
+2{\rm Re}(F_{A0}^\prime {\tilde F}_{V}^{\prime*})f_8(x,y)\nonumber\\&&
+2{\rm Re}(F_{A0}^\prime g^{T*}_{RR}F_T^*)f_9(x,y)
+2{\rm Re}({\tilde F}_{A+} g^{T*}_{RR}F_T^*)f_{10}(x,y)\nonumber\\&&
+2{\rm Re}({\tilde F}_{A-}^\prime g^{T*}_{RR}F_T^*)f_{11}(x,y)
+2{\rm Re}(F_{V}^\prime g^{T*}_{RR}F_T^*)f_{12}(x,y)~,
\end{eqnarray}
where we have neglected the term proportional to $|g^T_{RR}|^2$.
The functions $f_i(x,y)$ are also given in the appendix.
As in the case of $B\to D\ell\bar{\nu}$, $\rho_{D^*}(x,y)$ is related to 
the partial decay rate as
\begin{equation}
\frac{d^2\Gamma(B\to D^* \ell\bar{\nu})}{dxdy}
= \frac{G_F^2|V_{cb}|^2m_B^5}{128\pi^3}\rho_{D^*}(x,y)~.
\end{equation}
Keeping only ${\cal O}(g)$, the first three terms of Eq.(\ref{pol2}) become
\begin{eqnarray}
{\rm Im}(F^\prime_{A0}{\tilde F}^{*}_{A+})&=&
 4{\rm Im}(g^T_{RR})\frac{F_TF_{A+}}{\sqrt{r_\ell}}
 \Bigg\{\Bigg(1+\frac{1}{\sqrt{r_{D^*}}}\Bigg)
  \Bigg(1-\frac{x}{2}-y-r_\ell\Bigg)-r_\ell\Bigg\}\nonumber\\
{\rm Im}(F^\prime_{A0}{\tilde F}^{\prime*}_{A-})&=&
 {\rm Im}(G_P)\frac{F_{A0}F_{A-}}{\sqrt{r_\ell}(\sqrt{r_b}+\sqrt{r_c})}
 \Bigg[\frac{F_{A0}}{F_{A-}}+\frac{F_{A+}}{F_{A-}}(1-r_{D^*})+1+r_{D^*}-x\Bigg]
\nonumber\\&&
 +{\rm Im}(g^T_{RR})\frac{4F_{A0}F_T}{\sqrt{r_\ell}}\Bigg[
 1+\frac{F_{A-}}{F_{A0}}\Bigg\{\Bigg(1+\frac{1}{\sqrt{r_{D^*}}}\Bigg)
 \Bigg(1-\frac{x}{2}-y-r_\ell\Bigg)-r_\ell\Bigg\}\Bigg]
\nonumber\\
{\rm Im}({\tilde F}_{A+}{\tilde F}^{\prime*}_{A-})&=&
 {\rm Im}(G_P)\frac{F_{A+}F_{A-}}{\sqrt{r_\ell}(\sqrt{r_b}+\sqrt{r_c})}
 \Bigg[\frac{F_{A0}}{F_{A-}}+\frac{F_{A+}}{F_{A-}}(1-r_{D^*})+1+r_{D^*}-x\Bigg]
 \nonumber\\&&
 +{\rm Im}(g^T_{RR})\frac{4F_{A+}F_T}{\sqrt{r_\ell}}~.
 \label{imf}
\end{eqnarray}
The remaining four terms are all proportional to 
$\sim{\rm Im}(F^\prime({\rm or}~{\tilde F})g^T_{RR})\simeq F{\rm Im}(g^T_{RR})$
up to the linear order of $g$.
\par
The average polarization over the whole phase space is a convenient concept
because it measures the difference between the lepton numbers with opposite
transverse polarization to the decay plane divided by the total number of
leptons.
It is given by
\begin{equation}
\overline{P^\perp_{D^{(*)}}}=
 \frac{\int dxdy \rho_{D^{(*)}}(x,y)P^\perp_{D^{(*)}}(x,y)}
 {\int dxdy \rho_{D^{(*)}}(x,y)}~.
\end{equation}
In models where the couplings are proportional to the lepton mass such as 
multi-Higgs-doublet models and $R$-parity conserving
SUSY models, the polarization is proportional to the lepton mass.
In these cases the transverse polarization is large if the lepton is $\tau$.
When doing the numerical analysis, we only consider the $\tau$ production and
the results are summarized in Table 1.

\section{Results and Discussions}

As mentioned earlier in the Introduction, the main uncertainty comes from the 
hadronic form factors, or the IW function $\xi(\omega)$.
It needs nonperturbative methods to see its $\omega$ dependence.
We adopt two kinds of IW functions in the analysis;
$\xi(w)=1-0.75(w-1)$ from \cite{Wu}, 
$\xi(w)=1-1.13(w-1)$ from QCD sum rule \cite{Neubert}.
From Table 1, it seems that the structure of IW function does not affect 
$\overline{P^\perp_{D^{(*)}}}$ significantly.
The reason is that the kinematically allowed range of $\omega$ is quite narrow;
$1\le\omega\le(m_B^2+m_{D^{(*)}}^2)/(2m_B m_{D{(*)}})\simeq 1.59$, while the
interceptions and the slope parameters are not so far apart.
\par
One thing to be noticed in Table 1 is that 
the tensor interaction effects appear only through $g^T_{RR}$ multiplied by the
factor of $F_T$ in Eq.\ (\ref{FT}).
This means that only right-handed $b$-quark involves the tensor effect.
And $B\to Dl\bar{\nu}$ decay is more sensitive to ${\rm Im}g^T_{RR}$. 
Tensor contribution is almost ten times larger in $B\to Dl\bar{\nu}$ than
in $B\to D^*l\bar{\nu}$.
Since the kinematical factors of the terms proportional to ${\rm Im}g^T_{RR}$ 
are not suppressed compared to those of scalar or pseudoscalar couplings, 
various contributions of $g^T_{RR}$ in Eq.\ (\ref{pol2}) are destructive.
\par
\begin{figure}
\begin{center}
\epsfig{file=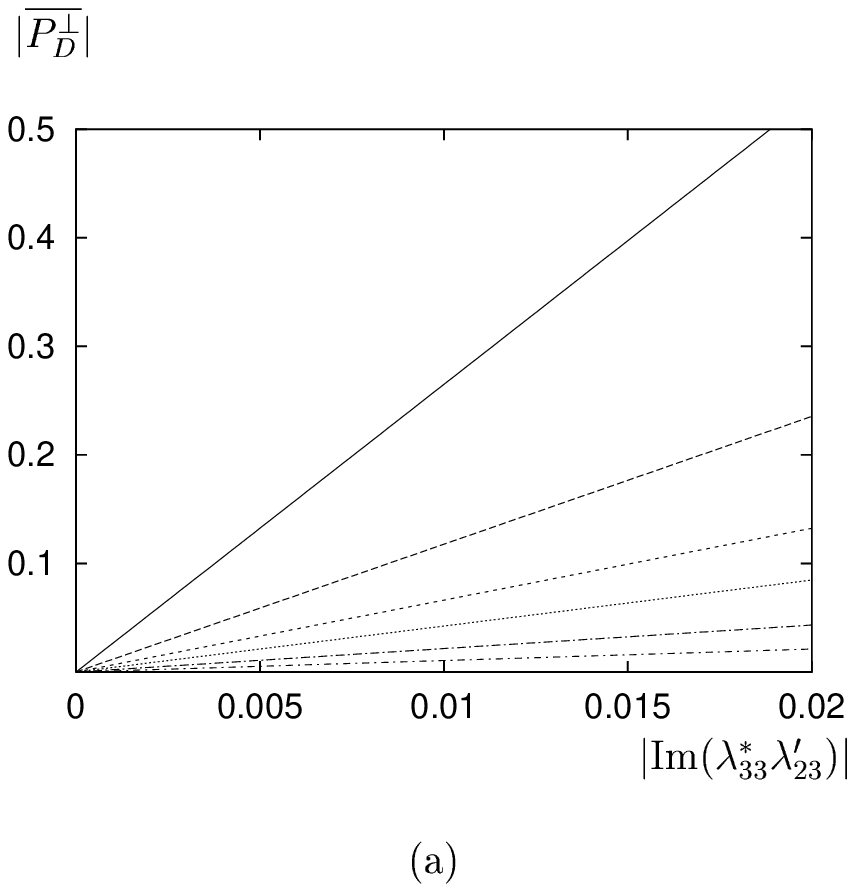}
\epsfig{file=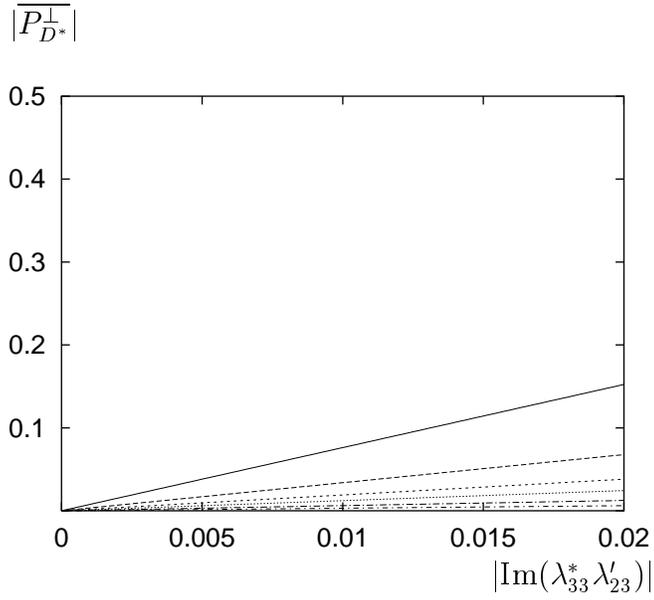}
\end{center}
\caption{Plots of $|\overline{P^\perp_{D^{(*)}}}|$ as a function of 
$|{\rm Im}(\lambda^*_{33}\lambda'_{23})|$ in the scalar leptoquark model.
Each line corresponds to $m_\phi=200$, $300$, $400$, $500$, $700$, $1000$ GeV
from the top, respectively.
In this Figure, we fix $\xi(\omega)=1-1.13(\omega-1)$.}
\label{Fig1}
\end{figure}

Note that our results are model independent and the model application is quite
straightforward.
Among the various extensions of the SM, the leptoquark model is a good candidate
to test the possible tensor interactions \cite{LQ}.
Leptoquarks are coupled to the lepton-quark pair.
As an example, consider only the scalar leptoquark $\phi$ which interacts with
quarks and leptons via the following Lagrangian:
\begin{equation}
{\cal L}_{LQ}=(\lambda_{ij}\bar{Q}_i e_{Rj}+\lambda'_{ij}\bar{u}_{Ri}L_i)\phi
+{\rm h.c.}~,
\end{equation}
where $Q$ and $L$ are quark and lepton doublets respectively, 
$\lambda_{ij}^{(')}$ are the couplings, and $i,j$ are the family indices.
After the Fierz reordering, the effective four-fermion interaction involving 
$\phi$ is described by (considering only $\tau$ lepton)
\begin{eqnarray}
{\cal L}_{\rm eff}&=&-\frac{1}{2}\frac{\lambda^*_{33}\lambda'_{23}}{m_\phi^2}
\Bigg[(\bar{c}_Rb_L)(\bar{\tau}_R\nu_{\tau L})
+\frac{1}{4}(\bar{c}_R\sigma_{\mu\nu}b_L)(\bar{\tau}_R\sigma^{\mu\nu}\nu_{\tau L})
+\frac{1}{8}(\bar{c}_L\sigma_{\mu\nu}b_R)(\bar{\tau}_R\sigma^{\mu\nu}\nu_{\tau L})\nonumber\\
&&+\frac{1}{8}(\bar{c}_R\sigma_{\mu\nu}b_L)(\bar{\tau}_L\sigma^{\mu\nu}\nu_{\tau R})
\Bigg]~.
\end{eqnarray}
Comparing with Eq.\ (\ref{new}), 
\begin{eqnarray}
g^S_{LR}&=&-\frac{\sqrt{2}}{G_FV_{cb}}
  \frac{\lambda^*_{33}\lambda'_{23}}{2m_\phi^2}~,~~~
g^S_{RR}=0~,\nonumber\\
g^T_{LR}&=&\frac{1}{4}g^S_{LR}~,~~~
g^T_{RR}=g^T_{LL}=\frac{1}{8}g^S_{LR}~.
\end{eqnarray}
With the typical values of $m_\phi=200$ GeV and
$|{\rm Im}(\lambda^*_{33}\lambda'_{23})|=0.01$ \cite{LQ,cskim}, we have
$|\overline{P^\perp_D}|\simeq 0.26$ and
$|\overline{P^\perp_{D^*}}|\simeq 0.076$.
Note that the different IW functions in Table 1 give almost the same 
value.
Figure \ref{Fig1} shows the model-parameter dependence of 
$|\overline{P^\perp_{D^{(*)}}}|$.
If the leptoquark mass goes beyond $\gtrsim 500$ GeV while retaining 
$|{\rm Im}(\lambda^*_{33}\lambda'_{23})|=0.01$, even $|\overline{P^\perp_D}|$
falls down to a few percent or less.
According to the above estimations, the observation of nonzero 
$|\overline{P^\perp_{D^(*)}}|$ will not only provide the new physics signals,
but also extract the tensor contributions.
A combined analysis of experimentally measured $\overline{P^\perp_D}$ and 
$\overline{P^\perp_{D^*}}$ will predict, in the leptoquark scenario,
$|\overline{P^\perp_D}/0.92(0.94)+\overline{P^\perp_{D^*}}/0.19|
\sim|{\rm Im}g^T_{RR}|$.
In the earlier work of \cite{cskim}, the optimal asymmetry of $B_{\ell 4}$
decay in the scalar leptoquark model is expected to be a good observable of CP
violation.
We argue that the analysis of lepton polarization given in this work will 
provide more chances to see new physics, especially tensor interactions.
\par
As a final remark, it should be noticed that the new physics effects can be 
nonzero, i.e., ${\rm Im}g_{\rm new}\neq 0$ even in the case $P^\perp=0$.
This is a new result of including tensor interactions.
Vanishing $P^\perp$ would constrain the involved couplings, giving a simple
relation between them.
We should, therefore, be cautious not to conclude that there is no signal of
new physics if $P^\perp_{D^{(*)}}=0$.

\section{Summary}

We give a model-independent analysis of transverse lepton polarization in 
exclusive $B\to D^{(*)}\ell\bar{\nu}$ decay including possible tensor 
interactions at the leading order of $1/m_Q$.
The results can directly be applied to specific models.
The transverse lepton polarization $P^\perp_{D^{(*)}}$ 
is a CP-odd observable and in general is
proportional to the imaginary part of the involved couplings.
Since in the SM the couplings are all real, $P^\perp_{D^{(*)}}$ is a 
good probe to observe the CP violation through the new physics.
In the leptoquark model, both of the scalar and tensor interactions contribute
to $P^\perp_{D^{(*)}}$, yielding 
$|\overline{P^\perp_D}|\simeq 0.26$ and
$|\overline{P^\perp_{D^*}}|\simeq 0.076$.
We find that in the leptoquark model, the tensor coupling is eight times 
smaller than the scalar one, and the effects of the tensor interactions can be 
extracted from the combined analysis of 
$\overline{P^\perp_D}$ and $\overline{P^\perp_{D^*}}$.

\begin{center}
{\large\bf Acknowledgments}\\[10mm]
\end{center}\par
This work was supported by the BK21 Program of the Korea Ministry of Education.

\begin{appendix}
\section{Kinematical Functions}

In this appendix we give the kinematical funcitions $g_i(x,y)$ and
$f_i(x,y)$. They are given by
\begin{eqnarray}
g_1(x,y)&=&(3-x-2y+r_\ell-r_D)(x+2y-1-r_\ell-r_D)\nonumber\\
   &&-(1+x+r_D)(1-x+r_D-r_\ell)~,\nonumber\\
   g_2(x,y)&=&r_\ell(3-x-2y-r_D+r_\ell)~,\nonumber\\
   g_3(x)&=&r_\ell(1-x+r_D-r_\ell)~,
\end{eqnarray}
and
\begin{eqnarray}
f_1(x,y)&=&(1-x+r_{D^*}-r_\ell)+\frac{1}{\sqrt{r_{D^*}}}
  (x+y-1-r_{D^*}-r_\ell)(1-y+r_\ell-r_{D^*})~,\nonumber\\
f_2(x,y)&=&[(x+2y-1-r_{D^*}-r_\ell)(3-x-2y-r_{D^*}+r_\ell)\nonumber\\&&
   -(1-x+r_{D^*}-r_\ell)
   (1+x+r_{D^*})]\Bigg(\frac{x^2}{4r_{D^*}}-1\Bigg)~,\nonumber\\
f_3(x,y)&=&r_\ell(1-x+r_{D^*}-r_\ell)\Bigg(\frac{x^2}{4r_{D^*}}-1\Bigg)~,\nonumber\\
f_4(x,y)&=&2xy(1-y+r_\ell-r_{D^*})+2x(2-x-y)(x+y-1-r_{D^*}-r_\ell)\nonumber\\&&
   -4(1-y+r_\ell-r_{D^*})(x+y-1-r_{D^*}-r_\ell)-4r_{D^*}y(2-x-y)~,\nonumber\\
f_5(x,y)&=&\frac{1}{\sqrt{r_{D^*}}}x(1-y)(x+y-1)
   -\frac{r_\ell}{2r_{D^*}}x(3-2x-3y-r_{D^*}+r_\ell)\nonumber\\&&
  +2(1-y)(1-x-y)-x+2r_{D^*}-r_\ell(x+y)~,\nonumber\\
f_6(x,y)&=&\frac{r_\ell}{2r_{D^*}}[x(1-y+r_\ell-r_{D^*})-2r_{D^*}(2-x-y)]
    ~,\nonumber\\
f_7(x,y)&=&r_\ell(3-x-2y-r_{D^*}+r_\ell)
   \Bigg(\frac{x^2}{4r_{D^*}}-1\Bigg)~,\nonumber\\
f_8(x,y)&=&2y(1-y+r_\ell-r{D^*})-2(2-x-y)(x+y-1-r_{D^*}-r_\ell)~,\nonumber\\
f_9(x,y)&=&
  -2\frac{1}{r_{D^*}\sqrt{r_\ell}}(1+1/\sqrt{r_{D^*}})
  (y-1)(x+y-1)(x+2y-2)\nonumber\\&&
  -2\frac{1}{\sqrt{r_{D^*}r_\ell}}(x+2y-2)(1+3r_\ell^2/r_{D^*}+r_\ell^2/\sqrt{r_{D^*}}
  +\sqrt{r_{D^*}})\nonumber\\
  &&+\sqrt{r_{D^*}r_\ell}[\{-8r_\ell/r_{D^*}+(2x+8y-4)/r_{D^*}\nonumber\\&&
  +4r_\ell^2/r_{D^*}^2
 (2x^2+12xy-12x+y^2-24y+12)/r_{D^*}^2+4\}]\nonumber\\&&
  +\sqrt{r_\ell}[(2x^2+8xy-8x+8y^2-16y+8)/r_{D^*}+6x+8y-12]~,\nonumber\\
f_{10}(x,y)&=&\sqrt{r_\ell r_{D^*}}\Big[
  4(y-1)r_\ell/r_{D^*}+2\{x(2y-3)+2(y-1)^2\}(x+2y-2)/(r_\ell r_{D^*})\nonumber\\&&
  (-3x^2-12xy+16x-12y^2+24y-12)/r_{D^*}\nonumber\\&&
  +2r_\ell^2x/r_{D^*}^2+r_\ell(-5x^2-8xy+8x)/r_{D^*}^2\nonumber\\&&
  -2x(y-1)(x+y-1)(x+2y-2)/(r_{D^*}^2r_\ell)\nonumber\\&&
  +x\{x^2+(x+10y-10)(x+y-1)\}/r_{D^*}^2
  +(4x-8y-8)/r_\ell-2x-4y+4\Big]\nonumber\\&&
  +\sqrt{r_\ell}\Big[
  r_{D^*}(4x-8y-8)/r_\ell+4r_{D^*}-r_\ell x(x-2y)/r_{D^*}\nonumber\\&&
  -2x(y-1)(x+y-1)(x+2y-2)/(r_{D^*}r_\ell)\nonumber\\&&
  +x\{x(5y-3)+2(y-1)(3y-2)\}/r_{D^*}
  +2\{x(2y-3)+2(y-1)^2\}\nonumber\\&&
  \times(x+2y-2)/r_\ell
  -3x^2-6xy+12x-4y^2+12y-12\Big]~,\nonumber\\
f_{11}(x,y)&=&\sqrt{r_\ell r_{D^*}}\Big[
  -4r_\ell(y-1)/r_{D^*}-(x+2y-2)(x-2y+2)/r_{D^*}\nonumber\\&&
  -2xr_\ell^2/r_{D^*}^2+r_\ell x(x+4y-4)/r_{D^*}^2\nonumber\\&&
  -x(y-1)\{x(y+1)+2(y-1)\}/r_{D^*}^2+2x+4y-4\Big]\nonumber\\&&
  +\sqrt{r_\ell}\Big[
  -4r_{D^*}+r_\ell x(x+2y-4)/r_{D^*}
  -x\{x^2+(x+y-1)(x+2y-4)\}/r_{D^*}\nonumber\\&&
  +4r_\ell+3x^2+6xy-4x+4y^2-12y+4\Big]~,\nonumber\\
f_{12}(x,y)&=&\frac{8}{\sqrt{r_\ell}}[
  (x-1)(y-1)(x+y-1)-(r_\ell-r_{D^*})^2 \nonumber\\&&
  -r_\ell(xy-2x-2y+2)(1+r_{D^*}/r_\ell)-r_\ell(x^2+r_{D^*}y^2/r_\ell)]~.
\end{eqnarray}

\end{appendix}


\newpage
%
%
\begin{table}
\vskip 5mm
\begin{center}
\begin{tabular}{c||ccc}
$\xi(w)$ & $1-0.75(w-1)$ \cite{Wu} & 
$1-1.13(w-1)$ \cite{Neubert} & 
In leptoquark model \\\hline
$\overline{P^\perp_D}$ &
$-0.92({\rm Im}G_S -2.2{\rm Im}g^T_{RR})$ &
$-0.94({\rm Im}G_S -2.2{\rm Im}g^T_{RR})$ &
0.26\\
$\overline{P^\perp_{D^*}}$ &
$-0.19({\rm Im}G_P -0.24{\rm Im}g^T_{RR})$ &
$-0.19({\rm Im}G_P -0.25{\rm Im}g^T_{RR})$ &
-0.076
\end{tabular}
\end{center}
\caption{
Numerical results of the average transverse lepton polarization in 
$B\to D^{(*)}\ell\bar{\nu}$
for different kinds of Isgur-Wise functions.
Estimations from the leptoquark model with $m_\phi=200$ GeV, 
$|{\rm Im}(\lambda^*_{33}\lambda'_{23})|=0.01$ are also given.}
\label{table1}
\end{table}


%
%
\end{document}